\documentclass[12pt,reqno]{amsart}
\usepackage{graphicx}
\usepackage{amssymb,amscd,amsbsy}
\usepackage{amsthm}

\usepackage{amsmath}
\usepackage{amsfonts}   
\usepackage{amssymb} 

\usepackage{graphicx}
\usepackage{amsmath}

\newcommand{\noi}{\noindent}

\def\X{\mathcal  X}

\def\L{\mathcal   L}
\def\M{\mathcal   M}
\def\A{\mathcal   A}
\def\B{\mathcal   B}
\def\cc{\mathfrak c}
\def\[{\left[}
\def\]{\right]}
\def\({\left(}
\def\){\right)}

\newcommand{\eeq}{\end{equation}}
\newcommand{\beq}{\begin{equation}}
\newcommand{\bay}{\begin{eqnarray}}
\newcommand{\ey}{\end{eqnarray}}
\newcommand{\bey}{\begin{eqnarray*}}
\newcommand{\eey}{\end{eqnarray*}}

\newtheorem{thm}{\hspace{\parindent}Theorem}[section]

\newtheorem{lem}[thm]{\hspace{\parindent}Lemma}

\theoremstyle{remark}
\newtheorem{rem}[thm]{Remark}
\newtheorem*{rem*}{Remark}

\usepackage{setspace}

\newcommand\epigraph[3]{
\vspace{1em}\hfill{}\begin{minipage}{#1}{\begin{spacing}{0.9}
\small\noindent\textit{#2}\end{spacing}
\vspace{1em}
\hfill{}{#3}}\vspace{2em}
\end{minipage}}

\begin{document}

\epigraph{2in}{''It is the theory that describes what we can observe''}{A. Einstein}
 


\newcommand{\vse}{\vspace{.2in}}
\numberwithin{equation}{section}

\title{INVESTOR'S SENTIMENT IN  MULTI-AGENT  MODEL OF THE CONTINUOUS DOUBLE AUCTION}
 
\author{A.Lykov, S.Muzychka and K.L.  Vaninsky}
\thanks{Corresponding author:   K.L. Vaninsky. vaninsky@math.msu.edu}
\begin{abstract}
We introduce and treat rigorously a new multi-agent  model of the continuous double auction or in other words the   order book (OB).  It is designed to explain  collective  behavior of the market when new information affecting the market  arrives. The novel feature  of the model  is two  additional slow changing parameters, the so--called sentiment functions. These sentiment functions  measure the conception of the fair price  of two groups of investors, namely,  bulls and bears. 
Our model specifies differential equations for the  time evolution of  sentiment functions   and constitutes a nonlinear Markov process  which  exhibits long term correlations.   We explain the intuition behind  equations for  sentiment functions and present numerical simulations which show that the behavior of our model is similar 
to the behavior of the real market. We also obtain a diffusion limit of the  model, the  Ornstein-Uhlenbeck type process with variable volatility. The volatility is proportional to  the difference of opinions  of bulls and bears about the fair price  of a security. 
The paper is complimentary to our previous work \cite{MV} where  mathematical proofs are presented.   
\end{abstract}
\maketitle

\tableofcontents 
\setcounter{section}{0}
\setcounter{equation}{0}

\section{Introduction. OB and volume.} 

\subsection{Non-equilibrium behavior of financial markets.}

Starting from Louis Jean-Baptiste Alphonse Bachelier,  people tried to model market behavior  using  various  stochastic processes.  Bachelier  himself in 1900 
used  Brownian motion \cite{Ba}. 
Subsequent attempts make  use of Markovian diffusions,  diffusions with jumps and even L\'{e}vy processes.  Models of  price change are  numerous, and  the  behavior of the market which  changes  its  statistics over time  is multi--faceted.   

In the last 15-20 years, all  US equity and futures exchanges have moved to an electronic order book and information about  active orders in the book is  available to all market participants.  
The term {\it market microstructure} was invented although  its definition varies depending on the author \cite{MM}. 

In  recent years, many  models of  market microstructure have been introduced and studied. Here we should  mention   papers \cite{Ro,  P}  in the financial literature on   various models of the book  and limit order markets,  and the papers  \cite{ AS,CST},  in which  stochastic  models of the order book  are  considered.   Since limit orders await  execution in FIFO queues,   these   models should be treated as  part of queuing theory.

The general Poisson framework  considered in our paper   \cite{GLV},  allows to simulate   various  real--life phenomena such as spread, V--shape of the order book, sudden price change under  diminishing liquidity in a book, {\it etc}.

In this paper,   we introduce and study a new multi--agent  nonlinear Markov model of the order book. It can be considered as an extreme  
degeneration of the model considered in \cite{GLV}.  We specify the behavior of  agents to model another phenomenon well known to traders in equity and futures  markets. 
The phenomenon  is depicted in Fig 1 (HLOC one minute bars),   which  contains  a  graph of the price of the index S\&P500 futures contract ESM08   and another graph of the trading volume  on Friday April 4  2008. This is the first Friday of the month and information about employment is released at 
8:30 am, \cite{Ya}. 
\newline
 
\centerline {\includegraphics[scale=0.4]{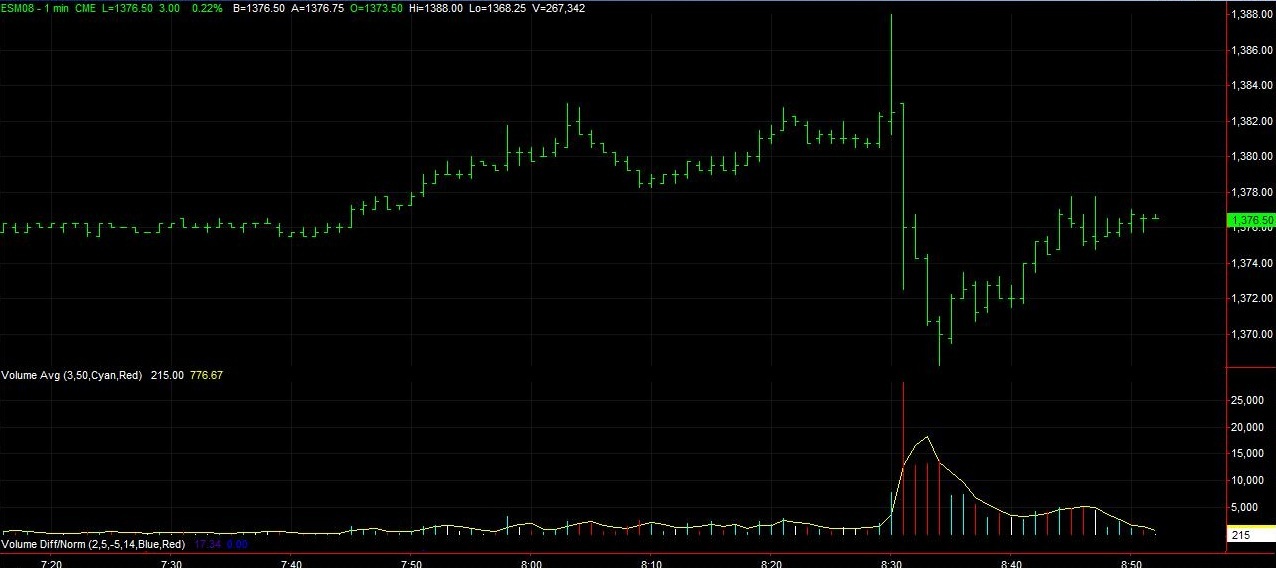}}
\centerline {Fig  1.}

\quad 

\noindent
In  the absence of any news before 8:30  the price changes in a diffuse manner and the volume of trading is fairly low.  
When  news hits the market,  the price  jumps under an avalanche of sell orders  and then slowly returns to the initial range. 
At the beginning of this process  the volume of trading increases significantly and then slowly decreases.  
This can be seen from the second graph  of volume of trades on Fig 1. Such mean--reverting behavior  is well known to intraday traders who follow the standard calendar of announcements of  economic indicators \cite{Ya}.

At the level of individual trades this phenomena is presented on Fig 2. As zero or starting moment  we have chosen the moment 8:30:00:000    
(hour:minute:sec: milsec). The horizontal axis represents a number of the trades counted from this moment of time. Each trade can be either on tick up or down. We construct two functions representing cumulative volume on tick up or down. The volume on tick up is represented by the yellow diagram. 
The volume on tick down is represented by the blue diagram.     
\newline
 
\centerline {\includegraphics[scale=0.4]{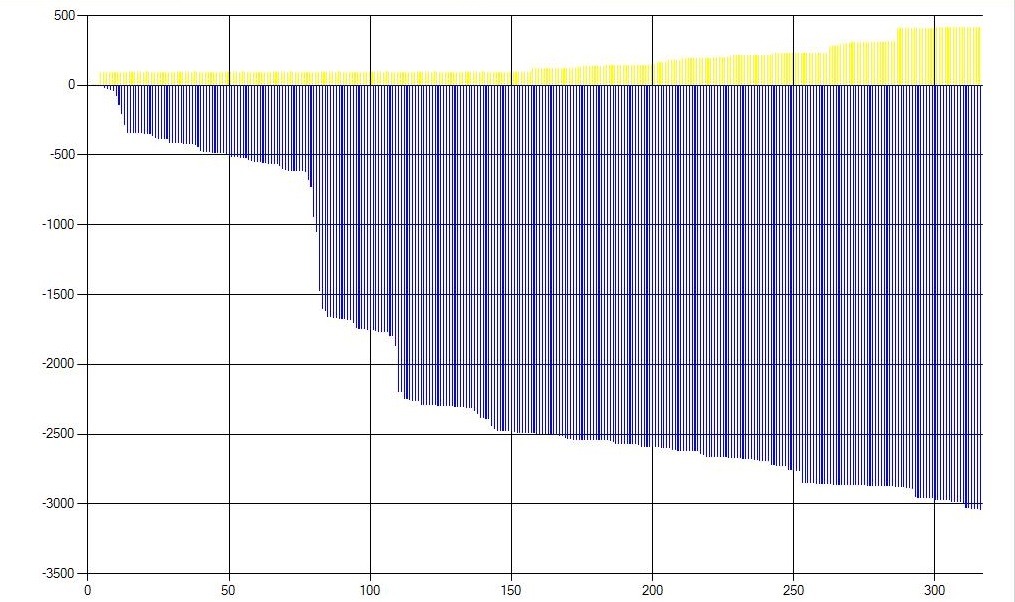}}
\centerline {Fig  2.}

\quad 

\noindent
In the first few seconds after the news announcement market participants submit about 300 orders to the exchange. They are mostly sell 
orders and their volume can be quite big.  This can be seen from  steep slope of the blue diagram. At the same time practically nobody submit buy orders to the exchange as can be seen from the yellow diagram. For this short period randomness disappears  and system behaves deterministically. 

To simulate this collective phenomenon  we propose a new,  completely rigorous mathematical model of the order book and price formation. Our model  is   based on ideas of nonequlibrium statistical mechanics. Proofs  of  the existence of the process and its convergence to an equilibrium  are given in a technical paper \cite{MV}.   All the mathematical results of this paper are new. Our approach is conceptually similar to an attempt to explain  classical   Brownian motion through a mechanical model of a heavy particle interacting with the  ideal gas, see \cite{SS}.  
 

The price changes are produced through the matching mechanism by the  interaction of  market participants with different roles. Previous models used ''zero--intelligence'' agents. Our agents are very slightly  more intelligent. 
They change the submission  rate  of  market orders to the matching mechanism in accordance with  certain slow variables    $L(t)$ and $M(t)$. 
Functions $L(t)$ and $M(t)$ measure the  market (bulls' and bears')  conception of the fair price of the security. They change relatively slowly compared to  the price,   which is the fast parameter.  In our approach,   the  collective mind of market participants transforms all public information available to it into two slowly changing functions $L$ and $M$
$$
\{\mbox{\it public  information available to traders }   \} 
\Longrightarrow \quad \{ L(t), \; M(t) \}. 
$$ 
If one could predict the behavior of the functions $L(t)$   and $M(t)$ for the real market then one could  have all 
money of the world.  Unfortunately at the moment, at least for the authors, this  is an  impossible task.

In order to simulate time evolution of the functions $L$ and $M$   we  use  ideas from   kinetic theory, {\it  i.e.}   ideas from the theory of the Vlasov equation. Originally this equation was written for plasma where ions interact with long range Coulomb forces  and their interaction  can not be neglected. The force acting on an ion can be computed by averaging the potential over the distribution of other ions  in the configuration space.  A mathematical model of this phenomenon was introduced by H.McKean in the form of a nonlinear Markov process \cite{Mc}.  In our model the  dynamics of parameters $L$ and $M$ at any moment of time  are   determined by the quantities  obtained by averaging over the distribution of price  at that moment. At a macro level,  the resulting  stochastic process obtained from a micro dynamics  is the discrete nonlinear analog of the classical  Ornstein-Uhlenbeck process.

Our  model   is analytically soluble and at the same time reproduces  the  complex behavior of the real market. 
The results of numerical simulations for our  model 
are  depicted  on  Fig 3.
\newline
 
\centerline {\includegraphics[scale=0.4]{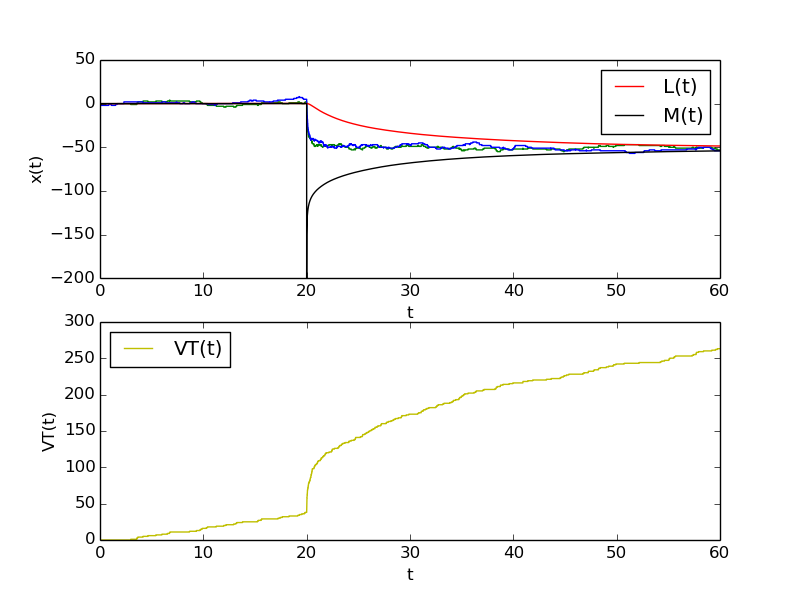}}
\centerline {Fig  3.}

\noindent 
The green and blue lines  on the upper graph represent randomly changing price in our model. We took two different samples of the process. 
The red and black  lines are  two slowly changing  functions $L(t)$ and $M(t)$.
The negative news hit the market at the moment 20 and the function  $M(t)$ drops  by 200.
The yellow  line $VT(t)$ on the lower graph represents  the cumulative volume of trading. Steep slope of the line for some time starting from the moment 20 is due to  increased volume of trading. 
These  graphs are  similar to their real counterparts on  Fig. 1.

The trade volume  at the level of micro-structure is similar to the real exchange. 
Cumulative aggregated volumes up $VU(t)$ and down $VD(t)$ on the time scale are given in Fig 4. 
The only difference here is the choice of the origin of the time scale. The news arrives now at the moment 20 
\newline
\centerline {\includegraphics[scale=0.4]{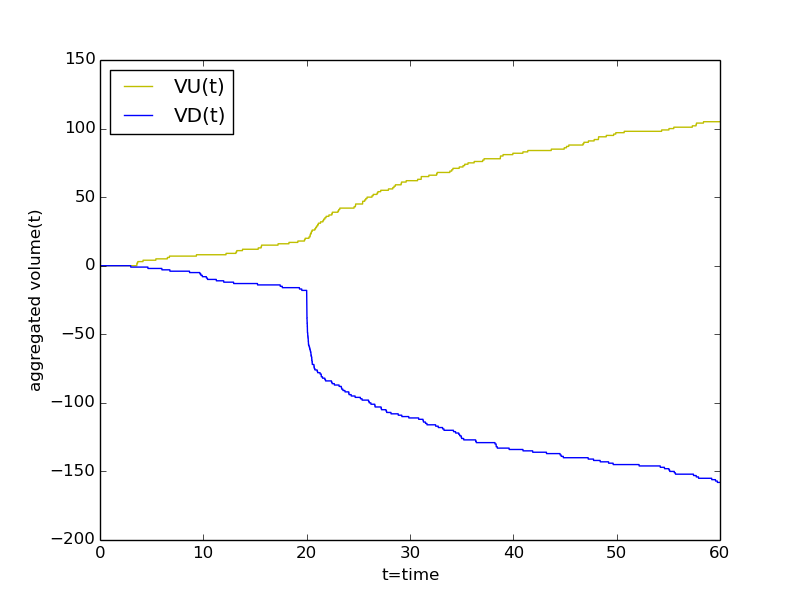}}
\centerline {Fig  4.}
\noindent 
The same cumulative  volumes $VU(n)$ and $VD(n)$    on the  trades scale are given in  Fig 5 
\newline
\centerline {\includegraphics[scale=0.4]{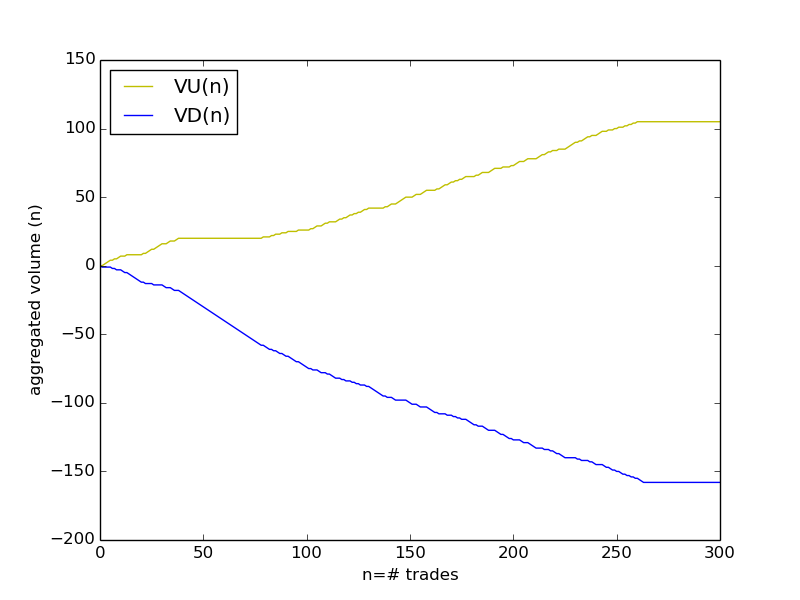}}
\centerline {Fig  5.}
\noindent 
The yellow graph of $VU(n)$ is flat right after the news arrival. It remains flat for some time  similar to its real counterpart in Fig 2. 
The blue graph of $VD(n)$ 
has a segment with the  steady slope $-1$ right after the news arrival meaning than only sell orders occur    for some period of time. 
This continues until the model reaches a new equilibrium and the graphs $VU(n)$ and $VD(n)$ stop to be  straight  lines. At this moment randomness is back to the system.

Now  we describe a   diffusion  limit of our discrete model. We believe that our variable volatility process is a  good candidate for derivative pricing.   
The diffusion limit is  a version of the classical OU process 
$\X(t)$  with  diffusion and  drift coefficients depending on the  (scaled) sentiment functions $\L(t)$ and $\M(t)$. 
It is   defined as  a solution of the stochastic differential equation 
$$
d \X(t)= e^{\cc(\L-\M)} \[dw(t) -  2 \cc \(\X(t)- \frac{\L+\M}{2}\) dt\], 
$$
where  $\cc >0$ and the functions $\L(t)$ and $\M(t)$ satisfy the system 
\begin{align}
	\frac{d\L(t)}{dt}=-e^{\cc(\L(t)-\M(t))}+1, \nonumber \\
	\frac{d\M(t)}{dt}=\;\;  e^{\cc(\L(t)-\M(t))}-1. \nonumber
\end{align}
As we see the process for the price of an asset is of Ornstein-Uhlenbeck type with variable volatility which is an exponent of the  
difference of opinions  $e^{\cc(\L-\M)}$. The equation for the difference   $\L-\M$  
$$
\frac{d(\L-\M)}{dt}=- 2\(  e^{\cc(\L-\M)}-1\) 
$$
can be solved explicitly.  Analysis of explicit formula (Section 5) shows that $\L(t)-\M(t)$  tends to zero as time goes to infinity. 

One can add jumps to
equations for $ \L(t) $ and $ \M(t) $ 
\begin{align}
	\frac{d\L(t)}{dt}=-e^{\cc(\L(t)-\M(t))}+1  + \sum_k {\A}_k\delta(t-\tau_k), \nonumber \\
	\frac{d\M(t)}{dt}=\;\;  e^{\cc(\L(t)-\M(t))}-1    - \sum_k {\B}_k\delta(t-\theta_k); \nonumber
\end{align} 
where $\A_k$   and $\B_k$ are positive random variables or  jumps which occur at the random moments of time  $\tau_k$ and $\theta_k$  and 
due to  the arrival of  new information.  
At the moments $\tau$'s and $\theta$'s the volatility changes    and then slowly tends to one as $\L(t)-\M(t)    $ tends to zero  under  the ODE dynamics. 

Starting from the work of Heston,  \cite{He},  models with random volatility play an important role in derivative pricing, see the paper \cite{FPS}. 
They all suffer from the apparent  drawback. Namely, what are  the origins of   variable volatility?  Within our approach drastic price changes   and variable volatility are attributed to the jump-like changes in the sentiment functions.

\subsection{Short description of relevant work.}

In the physics  and economics literature a large number   of  so--called multi-agent models were proposed recently. 
See  reviews   \cite{CTPA, SZSL, S,PGPS} and the original paper \cite{GS}.  
In these models,  idealized market participants or ''zero-intelligence''  agents submit  buy or sell orders (limit or market)  to the  matching mechanism,   
and these  orders are executed according to the rules of exchange. 

In the   recent paper \cite{BCVW} Brandouy, Corelli, Veryzhenko and Waldeck    ask whether ''zero--intelligence'' models produce a time series similar to  real markets. They claim that none of the  ''zero--intelligence'' models considered in their paper can    serve as a good approximation to   real--life phenomena. 
In our view   the  deficiency of  these models     is due to  stationarity in time  of statistical characteristics of  ''zero--intelligence'' agents. 
For example, these models are lucking an explanation of  sudden large price changes.  

The ideas that economic agents are not independent  and  that collective  phenomena are responsible for drastic changes in  market behavior are of course  not new.  There is a huge literature on what really  causes drastic changes. We should mention here a 
comprehensive survey article  of J.P. Bouchaud, J.D. Farmer and F. Lillo \cite{BFL}, and also surveys  \cite{BMP} and \cite{SFGK}. 
It is also well known  that there are various time scales in  time series or data  generated by financial markets.  There is also  an extensive  literature on  long term correlations in financial time series. For a   review see \cite{BFL,CL}.

Numerous academic publications in the field of behavioral finance  that try to analyze  investors' sentiment  using  Twitter \cite{BM}, or articles in Wall Street Journal or Dow Jones News or SEC filings  \cite{T} and \cite{CS}  have appeared recently. We have  to mention here an attempt  \cite{BFKR} to identify the news by the  type of information they provide  or  by tone.  There are also commercial companies  like Ravenpack which generate trading signals based on their proprietary  textual analysis systems. From this viewpoint our work is an attempt to incorporate sentiment as it is understood in behavioral finance   into a pricing/book mechanism.


\subsection{Content of the paper.}  A description of the matching mechanism and four groups of market participants are given in  Section 2, where  basic equations \ref{Kol} and \ref{mEq},  which define  the nonlinear Markov process are  introduced. To give the reader some intuition in Section 3 we consider the Ehrenfest model,   
the simplest stochastic model with mean--reverting behavior.  We study  in Section 4 equation (\ref{Kol}) defining  probabilities  of the fast variable $x(t)$. Equations (\ref{mEq}) defining the evolution of slow variables $L(t)$ and $M(t)$ are also considered there. The hydrodynamic limit is considered in Section 5. 
The continuum or diffusion limit is presented in Section 6. Finally, Section 7 contains results concerning  multi--particle approximation of the continuum system.  This multi--particles approximation is used for    numerical simulation of the continuum system. Appendix contains the proof of a technical result. 

\section{Description of the model.}

\subsection{Matching mechanism.} 
We now  describe  the matching mechanism. 
We adopt the following conventions. 
A queue is represented by a half--axis  as shown in Fig 6. Each order contains  a number of elementary units (contracts/stocks) and they are executed according to  
the  FIFO rule. Orders can be executed partially.   
\newline
 
\centerline {\includegraphics[scale=0.8]{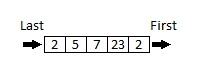}}
\centerline {Fig 6.}

\quad  

An order book is a structured list of interacting queues  as shown in  Fig 7.  Price levels are indexed by integers.  Each price level has two 
queues.   One queue contains orders to buy at a price not higher than this  price level; the second contains orders to sell at a price not lower than it.  
\newline
 
\centerline {\includegraphics[scale=0.8]{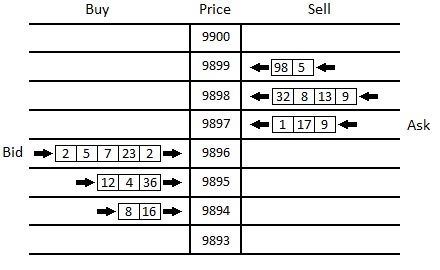}}
\centerline {Fig 7.}

\quad  
 
The book has this form due to the following rule. 
If some price level $X$ has $k$ orders to buy,   some price level $Y$ has $p$ orders to sell and  $X$ is greater or equal to $Y$,  
then $\min(k,p)$ orders are executed immediately  and removed from the system.  This is shown in Fig 8. 
\newline
 
\centerline {\includegraphics[scale=0.8]{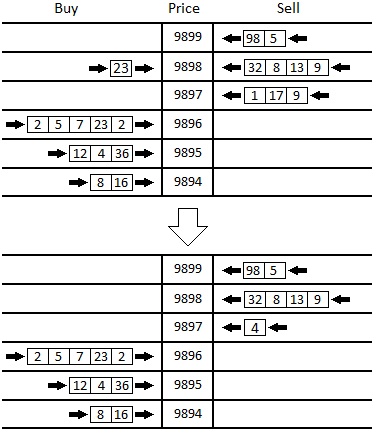}}
\centerline {Fig 8.}

\quad  
  
\noi
The real NYSE and NASDAQ books  show all orders at all levels. 
The   CME   book  shows only aggregated levels. 
\subsection{Market participants.} In the real market traders  submit market and limit  orders. The  behavior of market participants determines the state of the book and the evolution of the price. 

In our general framework   these trading activities are divided among  four groups of traders 
\begin{enumerate}
	\item Ask makers
	\item Bid makers
	\item Ask takers
	\item Bid takers 
\end{enumerate}

These participants can submit market orders of various sizes, limit orders and  can also  cancel existent limit orders in a book. Well--known facts about the spread between bid and ask, the V--shaped book profile and sudden price changes can be modeled within our framework.  Numerical experiments, \cite{GLV},  demonstrate that when 
market agents are calibrated according to the real data,  the results produced by the model are very similar to  empirical measurements. 

In this paper we want to model the  price change and the increase in  volume of trading  when news affecting the market arrives. In extremely liquid futures contract on the index S$\&$P  traded on CME   the spread is negligible  compared to the  price change (see Fig 1)  and the book is very dense at all ten visible price  levels closest to the current price. This justifies the assumptions  we make about the behavior of market agents (bid and ask makers) submitting limit orders to the book. 

In our model we assume that each group consists of one or a few traders.  This is a reasonable assumption  because traders are on a par  and we  consider the aggregated rate of  order submission  from an entire group. 
Ask/Bid makers fill each level of the book with just one contract  as shown in    Fig 9. 
\newline
 
\centerline {\includegraphics[scale=0.8]{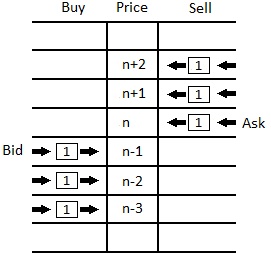}}
\centerline {Fig  9.}

\quad 

\noindent  In our model bulls correspond to  Ask takers and bears correspond to Bid takers. If an Ask taker submit buy order then she buys at the price $n$. If a Bid taker submits market a sell order then she sells at the price 
$n-1$. If an Ask/Bid taker sends the order and takes a contract on the  Ask/Bid then immediately the Bid/Ask maker fills the emptied level. Therefore the dynamics of the book are  determined by Ask/Bid takers. We consider the following stochastic process
$$ X(t)=(x(t),L(t),M(t)),$$ where $x(t)$ everywhere below denotes the level of ask at the moment $t$, and $L(t), M(t)$ are two slowly changing functions of time $t$ measuring what the market participants   take to be the  fair price of the security. Ask takers send the ''buy market'' orders with times between them being independent and exponentially distributed  at  the  rate 
$$\lambda_n(t)= e^{-c(n-L(t))}.$$ The number $c$ is some positive constant. When such an order arrives,  the level of ask changes from $n$ to $n+1$. 
Bid  takers   similarly  send  ''sell market''  orders  at the   rate 
$$\mu_n(t)= e^{c(n-M(t))}.$$ 
When this happens  the level of ask changes from $n$ to $n-1$.

Our model represents a huge simplification of the real order book. It has a much simpler phase space ($\mathbb{Z}$),  than the real book and therefore can be treated analytically. Literally,  it represents a price process, similar to a classical birth-death process. At the same time   it has a significant new feature; namely, its  jumps of nonstationary intensity determined by hidden functions  $(L(t),M(t))$  correspond to  the volume of trading.

The rates  produce an infinite system of Kolmogorov's equations for  probabilities
\beq\label{Kol}
\frac{\mathrm{d}p_n(t)}{\mathrm{d}t}
= \lambda_{n-1}(t)p_{n-1}(t) - (\lambda_n(t) + \mu_n(t))p_n(t) + \mu_{n+1}(t)p_{n+1}(t),\quad  n \in \mathbb{Z},
\eeq
and $p_{n}(t)=\mathrm{Prob}\{  x(t)=n \}$ for brevity. 
The functions $L(t)$ and $M(t)$ vary in time according to the differential equations
\begin{equation}
\begin{cases}
L'(t) & =-\sum_{n\in\mathbb{Z}}\lambda_{n}(t)p_{n}(t)+C_{\lambda} + \sum_k A_k\delta(t-\tau_k),\\
M'(t) & =\quad \sum_{n\in\mathbb{Z}}\mu_{n}(t)p_{n}(t)-C_{\mu} - \sum_k B_k\delta(t-\theta_k);
\label{mEq}
\end{cases}
\end{equation}
or equivalently
\begin{equation}
\begin{cases}
L'(t) & =-E\lambda_{x(t)}(t)+C_{\lambda} + \sum_k A_k\delta(t-\tau_k),\\
M'(t) & =\;\;\, E\mu_{x(t)}(t)-C_{\mu} - \sum_k B_k\delta(t-\theta_k).
\end{cases}\label{mEq1}
\end{equation}
Here $C_{\lambda}$ and $C_{\mu}$ are some positive constants. By default we assume $C_\lambda=C_\mu.$
The random times  $\tau$ and $\theta$  are Poisson flows and positive random amplitudes $A$ and $B$  are independent from them and distributed 
according to some probability law. 
We  call terms incorporated under the sum sign ''exterior forces''.

Note that $X(t)$ forms a {\it nonlinear Markov process}.  The transition probabilities of $x(t)$ depend on the distribution 
$p(t)=\{p_n(t)\}_{\mathbb{Z}}$ and the distribution of the exterior forces. 

Let us explain the intuition behind these equations.  In reality there are  two groups of  market participants,  bulls and bears.  Consider  bulls,  who believe that the price should be higher  and   submit buy  market orders  at the rate  $\lambda_n(t)$. Buying a contract  is an expense for them.  Bulls have two quantities in mind. One is the quantity  $C_\lambda$ which is how much they are willing to pay per infinitesimal unit of time, or in other words a rate of spending money. Another quantity is the function $L(t)$ which measures  what they believe  to be the  fair price. If  their conception of the  fair price is  too high then the quantity  $-E\lambda_{x(t)} + C_\lambda$ is negative and bulls decrease their expectations   according  to the  differential equations ($\ref{mEq}$).


When  good news affecting the market arrives, say at the moment $\tau_k$,  then the function $L(t)$  changes by a jump and $M(t)$ stays the same:  
$$
L(\tau_k-0) \longrightarrow L(\tau_k+0)=L(\tau_k-0)  +A_k,
$$
$$
M(\tau_k-0) \longrightarrow M(\tau_k+0)=M(\tau_k-0).
$$
A similar transformation occurs when  bad news arrives at some  moment $\theta_k$ 
$$
L(\tau_k-0) \longrightarrow L(\tau_k+0)=L(\tau_k-0) ,
$$
$$
M(\tau_k-0) \longrightarrow M(\tau_k+0)=M(\tau_k-0)-B_k.
$$
Since these transformations are very explicit and occur at  discrete 
moments of time we omit the news term in the  formulas below.    

The situation resembles  a random walk in a random environment. The functions $L(t)$ and $M(t)$ determine transition probabilities $P^{L,M}$ of $X(t)$ and expectations $E^{L,M}$ with respect to these probabilities. These functions are hidden parameters of the system and it is possible to   conditionalize  on them.   We conditionalize 
instead on the exterior forces $A$ and $B$ which, through differential equations,   determine the functions $L$ and $M$.   These  transition probabilities are denoted by $P^{A,B}$  and corresponding expectations are denoted by $E^{A,B}$. 
We refer to them as quenched probabilities and expectations. 
The Poisson measure  on  exterior forces $A$ and $B$ we denote by $P$. The product measure $P^{A,B}\times P$ denotes full transition probabilities 
of the process $X(t)$. The corresponding expectation we denote by $E_{X_t}$. 

Currently there is no  general theory of such processes. They are quite different from classical Markov processes. It can be shown that the process defined by 
\ref{Kol}--\ref{mEq} has a continuum of equilibrium states. It also possesses an integral of motion
$$
I= L(t)+ M(t) + \sum_{n\in \mathbb{Z}} p_n(t) n
$$
which does not change with time.   
  
Clearly the  behavior of idealized market participants could be   defined differently. For example, agents submitting limit orders can place  more than one order at each level. 
This assumption would allow to reproduce well-known, \cite{BMP,SFGK}, the V-shape distribution of the real book profile. Currently,  in such cases  
our  multiagent model  is  analyzed only numerically \cite{GLV}.

\section{The Ehrenfest model and continuous OU.} 

To give a reader some flavor of what is coming we start with the simplest  discrete analog  of the classical mean--reverting  Ornstein-Uhlenbeck process. 
In 1907 P. and T. Ehrenfest \cite{K} introduced a model which later became known as the "Ehrenfest urn model".  Fix an integer N and imagine two urns each containing a number of balls, in  such a way that the total number of balls in the two urns is 2N. At each moment of time we pick one ball at random (each with probability $1/2N$) 
and move it to the other urn. If $Y_t$ denotes the number of balls in the first urn minus $N$ then $Y_t,\;  t=0,1,2...;$ forms a Markov chain with the state space 
$\{-N,..., N\}$.  

This Markov chain is reversible with the  binomial distribution as a stationary measure
$$
\pi_k^{E}= \frac{2N!}{(N+k)! (N-k)!} \(\frac{1} {2}\)^{2N}, \qquad\qquad k = -N,..., N;
$$ 
and the generator
$$
\A^{E} f(k)=    \frac{1}{2}\( 1- \frac{k}{N}\) f(k+1) +  \frac{1}{2}\( 1+ \frac{k}{N}\) f(k-1) - f(k). 
$$

In 1930  Ornstein and Uhlenbeck  \cite{UO} introduced a model of  Brownian particle moving under  linear force. It is a Markov process with the state space $R^1$ and  the  generator  
$$
\A^{OU}= \frac {d^2}{d x^2} -2c x \frac{d }{d x}, \qquad\qquad c>0. 
$$
The OU  process is reversible with invariant density 
$$
\pi^{OU}(x)=\sqrt{\frac {c}{\pi}}\,  e^{-cx^2}.
$$

The two Markov processes are related. Under scaling 
\bey
   k &   \sim  &  x, \\
 {\rm spatial} \;\;\,   1 &   \sim  &  \frac{1}{\sqrt{n}}, \\
    N &   \sim &   \frac{\sqrt{n}}{c};  \\
\eey
the  generator $\A^{E}$ takes the form 
$$
\A^{E} f(x)=   \frac{1}{2} \( 1-  \frac{x c}{ \sqrt{n}}\) f(x+\frac{1}{\sqrt{n}}) +  \frac{1}{2} \( 1+ \frac{x c}{ \sqrt{n}}\) f(x-\frac{1}{\sqrt{n}})  - f(x).
$$
Using Taylor expansion after simple algebra we have 
$$
\A^{E} f(x)=\frac{1}{2n}\[  f''(x) -2c x f'(x)\] +...= \frac{1}{2n}\A^{OU} +... 
$$
This  shows on a formal level that $\A^{E}$converges to the generator $\A^{OU}$. Furthermore,  the     binomial   distribution $\pi_k^{E}$ converges to a Gaussian measure 
of   density  $\pi^{OU}(x)$.  Rigorous proof was obtained by M. Kac in \cite{KA} using  characteristic functions. 

\section{The discrete OU process. Speed and jump measure.}
\noi  
In this section we consider our model assuming $L$ and $M$ to be constant. The case when $L(t)$ and $M(t)$ satisfy ($\ref{mEq}$) we  consider  later.
 
We want to study the behavior of our  process: specifically, the probability of getting to infinity and the expected time  of  return  from infinity. 
Since both  spatial infinities are identical we consider the process on the right semi--axis.  A method of studying such processes 
by  means of an auxiliary space $(X,E)$ was introduced in a paper by W. Feller, \cite{F}. Let $$s = \frac{L+M}{2}.$$ The set $E$ consists of points 
$ \left\{ x_{n}\right\} _{n=0}^{\infty}\subset{\normalcolor \mbox{\ensuremath{\mathbb{R}_{\geqslant0}}}}$,  
 where 
 $$
 \text{\ensuremath{x_{0}=\frac{1}{\mu_{0}},}}
 $$
 $$
 x_{1}=x_{0}+\frac{1}{\lambda_{0}},
 \ldots
 $$
 $$
 x_{n+1}=x_{n}+\frac{\mu_{1}\mu_{2}\ldots\mu_{n}}{\lambda_{0}\lambda_{1}\ldots\lambda_{n}}=x_{n}+e^{cn(n+1)-2cns}\text{,}
 $$
 and 
 $$
 x_{\infty}=\lim_{n\rightarrow\infty}x_{n}=\infty.
 $$
 This fact implies that the process is recurrent; see  \cite{F},   section  16.b. 

The measure $ \mu$ is defined by the rule 
$$ 
\mu_{n}=\mu(\{x_{n}\})=\frac{\lambda_{0}\lambda_{1}\ldots\lambda_{n-1}}{\mu_{1}\mu_{2}\ldots\mu_{n}}=e^{-c(n-s)^{2}}.
$$
The ideal point  $x_{\infty}$  is an entrance point, {\it i.e.} the  expected  time of arrival  at  the finite part  of the  phase space starting at infinity is finite. 
This follows from the estimate 
$$
 {\textstyle {\textstyle {\displaystyle \sum_{n=0}^{\infty}}}}x_{n}\mu_{n} =\sum_{n=0}^{\infty}\sum_{k=0}^{n}e^{c(k(k-1)-2ks-(n-s)^{2})}
 \leqslant \mathrm{const}\cdot\sum_{n=0}^{\infty}(n+1)e^{-cn} < \infty .
$$
Whence,  $x(t)$ is very close to a Markov chain with a finite number of states. 

The detailed balance equations 
$$
\pi_n\lambda_n=\pi_{n+1} \mu_{n+1}
$$
are satisfied for the  distribution 
$$
\pi^s(n)=\frac{1}{\Xi}  e^{-c(n-s )^2}.
$$
The  normalization factor $\Xi=\Xi(s,c) $ is given by
$$
\Xi(s, c)= e^{-cs^2}   \Theta\( \frac{cs}{i \pi }, \frac{ci}{\pi}\),
$$ 
where the Jacobi theta function, \cite{HC}, is 
$$
\Theta(v,\tau)= \sum e^{2 \pi i v n + \pi i \tau n^2}.
$$  
It is interesting that the invariant measure $\pi$,  which should depend on both parameters $L$ and $M$,  depends on $s$ alone. 
The chain is ergodic and reversible.

For $L=M=0$ the  process $x_t$  has  a generator
$$
{\mathfrak A}  f(n)=  e^{-c n} f(n+1)  +  e^{c n}  f(n-1)  -  \(e^{-c n}   +  e^{c n}\)f(n).
$$
Let us show  how on a formal level $\A^{OU}$ can be obtained from the generator of the  discrete process. 
If one scales 
\bey
   n &   \sim  &  x, \\
 {\rm spatial} \;\;\,   1 &   \sim  &  \frac{1}{\sqrt{n}} ,\\
    c&   \sim &   \frac{c}{\sqrt{n}};  \\
 \eey
then 
$$
{\mathfrak A}  f(x)=  e^{-\frac{c}{\sqrt{n}} x} \[f(x+\frac{1}{\sqrt{n}}) -f(x)\]  +  
e^{\frac{c}{\sqrt{n}} x} \[ f(x-\frac{1}{\sqrt{n}})  -f(x) \]. 
$$
Using Taylor expansion  we have 
$$
{\mathfrak A}    f(x)= \frac{1}{n}  \[ f''(x)- 2cxf'(x)\]  +...=\frac{1}{n} \A^{OU}  f(x)+ .... 
$$
Apart from the factor $\frac{1}{n}$  this is a generator of the classical model. Furthermore,  under this scaling a discrete Gaussian type distribution $\pi^s(n)$ converges to the Gaussian distribution of   density $\pi^{OU}(x)$. It is natural to call   $x_t$  
a discrete  Ornstein-Uhlenbeck process.


In addition to $s$ it is natural to introduce another variable  
$$d= \frac{L-M}{2}.$$ The case when $d = 0$ we call agreement and $d \neq  0$ we call disagreement. The invariant measure does not depend on the parameter $d$ but the  
intensity of jumps does. Simple arguments show that at the equilibrium for any $d$
$$
E_{\pi^s} \lambda_{x(t)} = E_{\pi^s} \mu_{x(t)}.
$$
  In the  case of disagreement at the equilibrium {\it i.e.} $d\neq 0$,  
expectations are changed by  the exponential factor   $e^{cd}$
\beq\label{newexp}
E_{\pi^s} \lambda_{x(t)}^d=e^{cd} E_{\pi^s}\lambda_{x(t)}^{d=0},\qquad\qquad E_{\pi^s} \mu_{x(t)}^d=e^{cd} E_{\pi^s}\mu_{x(t)}^{d=0}.
\eeq
Apparently when $d>0$ the expectation increases and if  $d < 0$   it decreases. 

\section{The hydrodynamic limit.} 
\noi
We will now examine   various interesting limits of the  original system (\ref{Kol})--(\ref{mEq}). 
First, we will  exploit the multi--scale character of the dynamics and consider the so--called hydrodynamic limit of the system, see \cite{Var}. 

In this case one  assumes that the variables $L(t)$ and $M(t)$ change with macroscopic time $t$ while the process $x(\tau)$ moves with microscopic time $\tau=t/\epsilon$ where $\epsilon > 0.$ In the limit $\epsilon \rightarrow 0$ the process $x(\tau)$ is always at equilibrium {\it i.e.} it has 
stationary distribution $\pi^{s(t)}$ for all moments of time.  Therefore, using \ref{newexp}
\begin{align*}
	L'(t)= - ( e^{cd} - 1) E_{\pi^s} \lambda_{x(t)}  +\sum A_k \delta(t-\tau_k),\\
	M'(t) =+(  e^{cd} - 1) E_{\pi^s} \mu_{x(t)}    -    \sum   B_k \delta(t- \theta_k).
\end{align*}
Subtracting one equation from another and discarding the ''news'',  we obtain the following  equation for the function $d(t)$:
$$
d'(t)=-(e^{cd}-1)  V,  
$$ 
where $V= E_{\pi^s} \lambda_{x(t)}=E_{\pi^s} \mu_{x(t)}$.  This deterministic equation can be solved explicitly by the method of substitution. 

If a solution is negative  for some moment of time than it stays negative for all time and it is given by 
$$
d(t)= \frac{1}{c} \log \frac{ A e^{cVt}}{ A e^{cVt}+1 }. 
$$
The constant $A >0$ is determined by initial data. 
The solution increases monotonically  from negative infinity  to zero over the length of  the whole axis. For $A=V=1$ it is given below
\newline
 
\centerline {\includegraphics[scale=0.4]{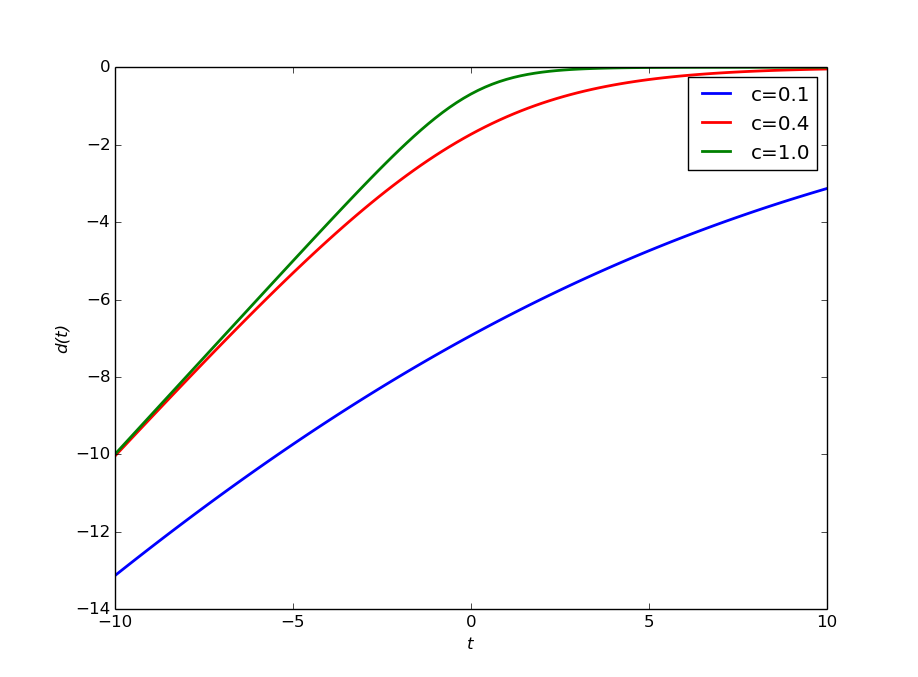}}
\centerline {Fig  10.}

\quad 

\noindent

In the opposite  case, if a solution is positive for some moment of time  then it  stays positive for all moments of time when it is  defined; namely for    $t > t_0= -\frac{1}{cV} \log A$. The solution is given by the formula 
$$
d(t) =\frac{1}{c} \log \frac{ A e^{cVt}}{A e^{cVt}-1}, \qquad \qquad A>0.
$$
The solution is defined and decreases monotonically  from positive infinity to zero for   $t > t_0$. For $A=V=1$ and $t_0=0$ it is given below
\newline
 
\centerline {\includegraphics[scale=0.4]{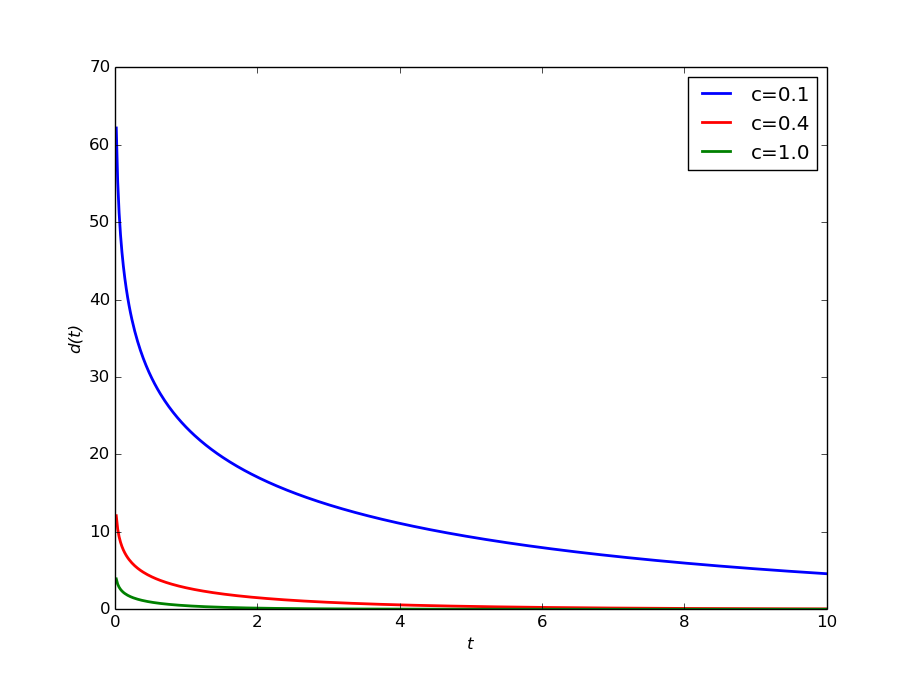}}
\centerline {Fig  11.}

\quad 

\noindent

\section{The continuum  limit.}
\noi
We now consider  a continuum limit of the discrete model. It is  instructive because  a version of the classical OU process $\X(t)$ arises with  diffusion and  drift coefficients depending on slow functions $\L(t)$ and $\M(t)$. 

The continuous time process $(\X(t),\L(t),\M(t))$ with values in $\mathbb{R}$
is   defined as  a solution of the stochastic differential equation 
$$
d \X(t)= e^{\cc(\L-\M)} \[dw(t) -  2 \cc \(\X(t)- \frac{\L+\M}{2}\) dt\], 
$$
where  $\cc >0$ and the functions $\L(t)$ and $\M(t)$ are
\begin{align}
	\frac{d\L(t)}{dt}=-e^{\cc(\L(t)-\M(t))}+1, \\
	\frac{d\M(t)}{dt}=\;\;  e^{\cc(\L(t)-\M(t))}-1. 
\end{align}
As we see the process for the price of an asset is of Ornstein-Uhlenbeck type with variable volatility which is an exponent of a  difference of opinions  $e^{\cc(\L-\M)}$. 
The equation for the difference of opinions  $\L-\M$  can be solved explicitly using substitution similar to the hydrodynamic limit.

There is a simple formal scaling  relation between the  two models.   
The  generator of the discrete model  has the form 
$$
{\mathfrak A}  f(n)=  e^{-c(n-L(t))} f(n+1)  +  e^{c(n-M(t))}  f(n-1)  -  \(e^{-c(n-L(t))}   +  e^{c(n-M(t))}\)f(n),
$$
or in the terms of $c$ and $d$
$$
{\mathfrak A}  f(n)= e^{cd}\[  e^{-c(n-s)} f(n+1)  +  e^{c(n-s)}  f(n-1)  -  \(e^{-c(n-s)}   +  e^{c(n-s)}\)f(n)\].
$$
If one scales 
\bey
   n &   \sim  &  x, \\
 {\rm spatial} \,   1 &   \sim  &  \frac{1}{\sqrt{n}}, \\
    c&   \sim &   \frac{\cc}{\sqrt{n}},  \\
   L &   \sim  &  \frac{\L+\M}{2} +\sqrt{n}(\L-\M),\\
    M &   \sim  &  \frac{\L+\M}{2} -  \sqrt{n}(\L-\M);\\
\eey
then 
$$
s \sim \frac{\L +\M}{2},  \qquad\qquad  d \sim \sqrt{n} (\L-\M); 
$$
and for the generator we have 
$$
{\mathfrak A}  f(x)\sim \frac{1}{n} e^{\cc(\L-\M)}  \[ f''(x)- 2 \cc(x- s)f'(x)\]  +... .  
$$
Apart from the factor $\frac{1}{n}$,   this is the  generator of the continuous model.  The time scaling $t \sim n t $ removes this factor. 
The differential equations for $L$ and $M$,  when they are  scaled the same way,   produce    differential equations for the functions $\L$ and $\M$.  
In fact, equations for $\L$ and $\M$  decouple  from the stochastic component $\X(t)$  and can be solved explicitly as in the hydrodynamic limit above. We do not dwell on this. 

If $\L(t)=\M(t) =s$,  then  $\X(t)$ is just a classical Ornstein-Uhlenbeck process with mean $s$. There is a proof similar to that given in  \cite{MV} that if the process starts with  some $\L(0)\neq \M(0)$,  then $\X(t)$  converges to OU and $\L(t)$ and $\M(t)$  converge to the same  constant $s$.

\section{Multiparticle approximation  and Monte-Carlo simulation.}
\noi
In this section for the purpose of numerical simulation we consider a multi-particle approximation of the  process   defined by \ref{Kol}--\ref{mEq}.
We assume that there is  no news in the model; that is to say,  $A_k=B_k=0.$ To simplify notations everywhere below we assume $c=1.$ Fix $N\in\mathbb{Z}$
and consider the following $N$-particle process 
$$
X^{N}(t)=(x_{1}^{N}(t),x_{2}^{N}(t),\ldots,x_{N}^{N}(t),L^{N}(t),M^{N}(t)), 
$$
where $x_{i}^N(t)\in\mathbb{Z},$ $i\in\left\{ 1,\ldots,N\right\} $
represents the coordinate of the $i$-th particle jumping on $\mathbb{Z}$
with intensities 
\begin{align*}
n\to n+1:\quad & \ \lambda_{n}^{N}(t)=e^{c(-n+L^{N}(t))},\\
n\to n-1:\quad & \ \mu_{n}^{N}(t)=e^{c(n-M^{N}(t))};
\end{align*}
 and $(L^{N}(t),M^{N}(t))\in\mathbb{R}^{2}.$ Denote by
$p^{N}(t)=\left\{ p_{n}^{N}(t)\right\}$
the empirical distribution of $(x_{1}^{N}(t),x_{2}^{N}(t),\ldots,x_{N}^{N}(t))$,
namely 
$$
p_{n}^{N}(t)=\frac{1}{N}\sum_{i=1}^{N}I_{\{x_{i}^{N}(t)=n\}},
$$
and suppose that $(L^{N}(t),M^{N}(t))$ satisfies the following equations:
\begin{equation}
\begin{cases}
\frac{\mathrm{d}}{\mathrm{d}t}L^{N}(t) & =-\sum_{n\in\mathbb{Z}}\lambda_{n}^{N}(t)p_{n}^{N}(t)+C_{\lambda},\\
\frac{\mathrm{d}}{\mathrm{d}t}M^{N}(t) & =\quad \sum_{n\in\mathbb{Z}}\mu_{n}^{N}(t)p_{n}^{N}(t)-C_{\mu};
\end{cases}\label{eq:ZN1}
\end{equation}
or equivalently 
\begin{equation}
\begin{cases}
\frac{\mathrm{d}}{\mathrm{d}t}L^{N}(t) & =-\frac{1}{N}\sum_{i=1}^{N}\lambda_{x_{i}^{N}(t)}^{N}(t)+C_{\lambda},\\
\frac{\mathrm{d}}{\mathrm{d}t}M^{N}(t) & =\quad\frac{1}{N}\sum_{i=1}^{N}\mu_{x_{i}^{N}(t)}^{N}(t)-C_{\mu}.
\end{cases}\label{eq:ZN2}
\end{equation}
Finally,  suppose that $(x_{1}^{N}(0),x_{2}^{N}(0),\ldots,x_{N}^{N}(0))$
are independent $p(0)$-distributed values,  and $L^{N}(0)=L(0),$ $M^{N}(0)=M(0),$
are some fixed values.
\begin{rem}
Note that while for each $n\in\mathbb{N},$ $N\in\mathbb{N}$ and
$t\in[0,\infty)$,  functions $p_{n}(t),$  $L(t)$ and $M(t)$ are deterministic,  and 
$p_{n}^{N}(t),$ $L^{N}(t)$ and $M^{N}(t)$ are random variables
(not deterministic).
\end{rem}
\noi
Our goal is to show that in some sense $X^{N}$ converges to $X$
as soon as $N\to\infty.$

Consider the following one-parametric series of Banach-spaces:
$$
B_{\alpha}=\left\{ x:\|x\|_{\alpha}<\infty\right\} ,\quad\|x\|_{\alpha}=\sum_{n\in\mathbb{Z}}\alpha_{n}|x_{n}|,\ \mathrm{where}\ \alpha_{n}=e^{\frac{cn^{2}}{2}+\alpha|n|},\ \alpha\in\mathbb{R}.
$$
The main result of \cite{MV} is the following theorem: 
\begin{thm}
\label{thm:existence}For any $\alpha\in\mathbb{R}$ and initial probability
measure $p(0)\in B_{\alpha}$,  the process $X(t)$ is correctly defined
on the half-line $[0,\infty)$. Moreover, $p(t)$ is a continuous
$B_{\alpha}$-valued function.
\end{thm}
Consider the  one-parametric series of Hilbert-spaces
$H_{\alpha},\ \alpha\in\mathbb{R}$ consisting of the infinite series
$\xi=\left\{ \xi_{n}\right\} _{n\in\mathbb{Z}}$ with inner product
$$
\langle\xi,\eta\rangle=\sum_{n\in\mathbb{Z}}\alpha_{n}\cdot\xi_{n}\eta_{n}
$$
 and denote by $|\cdot|_{\alpha}$ the corresponding norm ($|\xi|_{\alpha}^{2}=\langle\xi,\xi\rangle$).
We prove the following theorem:
\begin{thm}
\label{thm:convergence}Let $p(0)\in B_{\alpha+1},$ $\alpha\in\mathbb{R}.$
Assume that $E\|p^{N}(0)-p(0)\|_{\alpha+1}\to0$ as $N\to\infty.$
Then for all $t\in[0,\infty)$ 
$$
\sup_{s\leq t}E|p^{N}(s)-p(s)|_{\alpha}^{2}\to0\ as\ N\to\infty.
$$
\end{thm}
\noi
A proof of this theorem can be found in the  Appendix.

In order to produce the simulations  depicted  in  Fig 3, 4 and 5 we used the 
multiparticle approximation with $N=50$ particles $\left\{ \xi_{i}\right\} _{i=1}^{N}$
described above. 
All particles $\left\{ \xi_{i}\right\} $ jump with  the same rates
$$
\lambda_{N}(\xi_{i})=e^{-c(\xi_{i}-L_{N})},\;\;\quad\quad  \mu_{N}(\xi_{i})=e^{c(\xi_{i}-M_{N})}.
$$
We fix  $c=0.05$ and $C_{\lambda}=1.02$. 
There is a jump on Fig 3  at the moment $t=20$ of the event $M\to M-B,$ where $B=200.$ 


{\bf Acknowledgment.} We would like to thank Mark Kelbert, Yurii Sukhov, Vadim Malyshev, Mikhail Soloveitchik  and Alexander Glekin  for many helpful remarks. K.V. thanks Henry McKean and Raghu Varadhan for discussions. We are grateful to  James Doyle for help with editing. 

\vskip .2in
\noindent
Kirill Vaninsky
\newline
Department of Mathematics
\newline
Michigan State University
\newline
East Lansing, MI 48824
\newline
USA
\vskip 0.1in
\noindent
\newline
vaninsky@math.msu.edu

\vskip .2in
\noindent
Stepan Muzychka 
\newline
Faculty of Mathematics and Mechanics 
\newline
Moscow State University 
\newline 
Vorobjevy Gory
\newline
Moscow, Russia
\noindent
\newline
stepan.muzychka@gmail.com

\vskip .2in
\noindent
Alexander Lykov 
\newline
Faculty of Mathematics and Mechanics 
\newline
Moscow State University 
\newline 
Vorobjevy Gory
\newline
Moscow, Russia
\noindent
\newline
alekslyk@yandex.ru
\newpage

\centerline{\bf Appendix.} 
We start with some simple statements about the properties of the sequence $\alpha_n$ and the introduced norm $\|\cdot\|_\alpha.$ Notice two simple inequalities that we will need afterwards 
\begin{align}
 & \alpha_{n+1}e^{-n}\leq\mathrm{const}\cdot\alpha_{n},\label{eq:aux1}\\
 & \alpha_{n-1}e^{n}\leq\mathrm{const}\cdot\alpha_{n}.\label{eq:aux2}
\end{align}
Also note that if $x=\left\{ x_{k}\right\} _{k\in\mathbb{Z}}\in B_{\alpha+1}$, 
then 
\begin{equation}
\|x'\|_{\alpha}\leq\|x\|_{\alpha+1},\ \mathrm{where}\ x'=\left\{ x_{k}e^{|k|}\right\} _{k\in\mathbb{Z}}.\label{eq:aux3}
\end{equation}
And finally for any random variable $\xi$ such that $P(\xi\in\mathbb{Z})=1$
we have 
\begin{equation}
\|p_{\xi}\|_{\alpha}=Ee^{\frac{c\xi^{2}}{2}+\alpha|\xi|},\label{eq:aux4}
\end{equation}
where $p_{\xi}$ is the distribution of $\xi.$ 

For our purposes it will be convenient to decompose the infinitesimal
operator $H(t)$ corresponding to the process $x(t)$ 
$$
\left(\begin{array}{ccccc}
\ldots & \ldots & \ldots & \ldots & \ldots\\
\ldots & -I_{n-1}(t) & \mu_{n-1}(t) &  & \ldots\\
\ldots & \lambda_{n}(t) & -I_{n}(t) & \mu_{n}(t) & \ldots\\
\ldots &  & \lambda_{n+1}(t) & -I_{n+1}(t) & \ldots\\
\ldots & \ldots & \ldots & \ldots & \ldots
\end{array}\right)\ (\mathrm{where}\ I_{n}(t)=\lambda_{n}(t)+\mu_{n}(t))
$$
into the sum of the diagonal and off-diagonal parts
$$
H(t)=H_{0}(t)+V(t).
$$
 Also we consider the similar decomposition $x_{i}^{N}(t),\ i\in\left\{ 1,\ldots,N\right\} $
$$
H^{N}(t)=H_{0}^{N}(t)+V^{N}(t).
$$

\begin{rem}
Note that due to symmetry $H^{N}(t)$ does not depend on $i.$
\end{rem}

Everywhere below we will denote by $\mathrm{c}(\cdot)$ any nonnegative
nondecreasing function of some parameters. It will also be convenient
to introduce the following notation 
$$
\Delta^{N}(t)=p^{N}(t)-p(t).
$$
We need the following auxiliary lemma.
\begin{lem}
\label{lem:auxLem}For all $t\in[0,\infty)$ and $n\in\mathbb{N}$\end{lem}
\begin{enumerate}
\item $\ensuremath{e^{L(t)}\leq\mathrm{c}(t),\ }\ensuremath{e^{-M(t)}\leq\mathrm{c}(t),\ }\ensuremath{e^{L^{N}(t)}\leq\mathrm{c}(t),\ }\ensuremath{e^{-M^{N}(t)}\leq\mathrm{c}(t).}$
\item $\ensuremath{|e^{\pm L(t)}-e^{\pm L^{N}(t)}|\leq\mathrm{c}(t)\Gamma(t)}$
and $\ensuremath{|e^{\pm M(t)}-e^{\pm M^{N}(t)}|\leq\mathrm{c}(t)\Gamma(t)}$, 
where 
$$
\Gamma(t)=\int_{0}^{t}\sum_{n\in\mathbb{Z}}|\Delta_{n}^{N}(s)|e^{|n|}\mathrm{d}s.
$$

\item $\ensuremath{E\Gamma^{2}(t)\leq\mathrm{c}(t)\sup_{s\leq t}E|\Delta^{N}(s)|_{\alpha}^{2}.}$
\item If $\ensuremath{p(0)\in B_{\alpha},}$ then 
$$
\ensuremath{E|p(t)(H_{0}^{N}(t)-H_{0}(t))|_{\alpha}^{2}\leq\mathrm{c}(t)\sup_{s\leq t}E|\Delta^{N}(s)|_{\alpha}^{2}.}
$$

\item If $\ensuremath{p(0)\in B_{\alpha},}$ then 
$$
\ensuremath{E|p(t)(V^{N}(t)-V(t))|_{\alpha}^{2}\leq\mathrm{c}(t)\sup_{s\leq t}E|\Delta^{N}(s)|_{\alpha}^{2}.}
$$

\item If $p(0)\in B_{\alpha},$ then 
$$
\textup{\ensuremath{E\langle\Delta^{N}(t),\Delta^{N}(t)H^{N}(t)\rangle_{\alpha}\leq\mathrm{c}(t)\sup_{s\leq t}E|\Delta^{N}(s)|_{\alpha}^{2}.}}
$$

\item Let $\ensuremath{E\|p^{N}(0)\|_{\alpha+1}<\infty}$,  then 
$$
\sum_{n}\alpha_{n}EQ_{n}^{N}(t)\leq c(t,E\|p^{N}(0)\|_{\alpha+1}),
$$
where 
$$
Q_{n}^{N}(t)=p_{n-1}^{N}(t)\lambda_{n-1}^{N}(t)+p_{n+1}^{N}(t)\mu_{n+1}^{N}(t)+p_{n}^{N}(t)\lambda_{n}^{N}(t)+p_{n}^{N}(t)\mu_{n}^{N}(t).
$$
\end{enumerate}
We will now  prove  the main theorem. (The proof of the lemma is very technical and we prefer to present it later). During time $\mathrm{d}t$
the variable $p^{N}(t)$ can be changed in the following way: 
$$
p_{n}^{N}(t)\to\begin{cases}
p_{n}^{N}(t)-\frac{1}{N}, & \quad\mathrm{with\ probability\ }P_{n}^{out}\mathrm{d}t:=Np_{n}^{N}(t)(\lambda_{n}^{N}(t)+\mu_{n}^{N}(t))\mathrm{d}t;\\
p_{n}^{N}(t)+\frac{1}{N}, & \quad\mathrm{with\ probability\ }\ P_{n}^{in}\mathrm{d}t:=\\
 & \ \ \ =N\mathrm{d}t(p_{n-1}^{N}(t)\lambda_{n-1}^{N}(t)+p_{n+1}^{N}(t)\mu_{n+1}^{N}(t));\\
p_{n}^{N}(t), & \quad\mathrm{with\ probability\ }1-P_{n}^{out}\mathrm{d}t-P_{n}^{in}\mathrm{d}t.
\end{cases}
$$
At the same time $p(t)$ is deterministic 
\begin{align*}
 & p_{n}(t)\to p_{n}(t)+(p(t)H(t))_{n}\mathrm{d}t=\\
 & =p_{n}(t)+(\lambda_{n-1}p_{n-1}(t)-(\lambda_{n}p_{n}(t)+\mu_{n}p_{n}(t))+\mu_{n+1}p_{n+1}(t))\mathrm{d}t.
\end{align*}
So 
$$
(\Delta_{n}^{N})^{2}\to\begin{cases}
(\Delta_{n}^{N}-\frac{1}{N})^{2}+O(\mathrm{d}t) & \mathrm{with\ probability\ }P_{n}^{out}\mathrm{d}t;\\
(\Delta_{n}^{N}+\frac{1}{N})^{2}+O(\mathrm{d}t) & \mathrm{with\ probability\ }P_{n}^{in}\mathrm{d}t;\\
(\Delta_{n}^{N}-(p(t)H(t))_{n}\mathrm{d}t)^{2} & \mathrm{with\ probability\ }1-P_{n}^{out}\mathrm{d}t-P_{n}^{in}\mathrm{d}t.
\end{cases}
$$
Applying Markov property and opening brackets we get
\begin{align*}
 & \mathrm{d}E\left((\Delta_{n}^{N})^{2}|X_{N}(t)\right)=\\
 & =P_{n}^{out}\mathrm{d}t\times\left\{ -\frac{2\Delta_{n}^{N}}{N}+\frac{1}{N^{2}}\right\} +P_{n}^{in}\mathrm{d}t\times\left\{ \frac{2\Delta_{n}^{N}}{N}+\frac{1}{N^{2}}\right\} +\\
 & +(1-P_{n}^{out}\mathrm{d}t-P_{n}^{in}\mathrm{d}t)\times\left\{ -2\Delta_{n}^{N}\cdot(p(t)H(t))_{n}\mathrm{d}t\right\} =\\
 & =\frac{1}{N^{2}}\left(P_{n}^{out}+P_{n}^{in}\right)\mathrm{d}t+\frac{2}{N}\Delta_{n}^{N}\left(P_{n}^{in}-P_{n}^{out}\right)\mathrm{d}t-\\
 & -2\Delta_{n}^{N}\cdot(p(t)H(t))_{n}\mathrm{d}t.
\end{align*}
Therefore, 
\begin{align*}
\frac{\mathrm{d}}{\mathrm{d}t}E(\Delta_{n}^{N})^{2} & =\frac{1}{N^{2}}E\left(P_{n}^{out}+P_{n}^{in}\right)+\frac{2}{N}E\Delta_{n}^{N}\left(P_{n}^{in}-P_{n}^{out}\right)-2E\Delta_{n}^{N}\cdot(p(t)H(t))_{n}.
\end{align*}
Summing up by $n$ we have 
\begin{align*}
 & \frac{\mathrm{d}}{\mathrm{d}t}E|\Delta^{N}|_{\alpha}^{2}=\frac{\mathrm{d}}{\mathrm{d}t}\sum_{n\in\mathbb{Z}}\alpha_{n}E(\Delta_{n}^{N})^{2}=\\
 & =\frac{1}{N}E\sum_{n\in\mathbb{Z}}\alpha_{n}\left(p_{n}^{N}\lambda_{n}^{N}+p_{n}^{N}\mu_{n}^{N}+p_{n-1}^{N}\lambda_{n-1}^{N}+p_{n+1}^{N}\mu_{n+1}^{N}\right)+\\
 & +2E\sum_{n\in\mathbb{Z}}\alpha_{n}\Delta_{n}^{N}\left(p_{n-1}^{N}\lambda_{n-1}^{N}+p_{n+1}^{N}(t)\mu_{n+1}^{N}-p_{n}^{N}\lambda_{n}^{N}-p_{n}^{N}\mu_{n}^{N}\right)-\\
 & -2E\sum_{n\in\mathbb{Z}}\alpha_{n}\Delta_{n}^{N}\cdot(pH)_{n}=\\
 & =\frac{1}{N}EQ^{N}+2E\langle\Delta^{N},p^{N}H^{N}\rangle_{\alpha}-2E\langle\Delta^{N},pH\rangle_{\alpha}=\\
 & =\frac{1}{N}EQ^{N}+2E\langle\Delta^{N},(p^{N}-p)H^{N}\rangle_{\alpha}+2E\langle\Delta^{N},p(H^{N}-H)\rangle_{\alpha}=\\
 & =\frac{1}{N}EQ^{N}+2E\langle\Delta^{N},\Delta^{N}H^{N}\rangle_{\alpha}+2E\langle\Delta^{N},p(H_{0}^{N}-H_{0})\rangle_{\alpha}+2E\langle\Delta^{N},p(V^{N}-V)\rangle_{\alpha}.
\end{align*}
Applying Cauchy-Schwarz inequality we get 
\begin{align*}
\frac{\mathrm{d}}{\mathrm{d}t}E|\Delta^{N}|_{\alpha}^{2} & \leq\frac{1}{N}EQ^{N}+\\
 & +2\left(E|\Delta^{N}|_{\alpha}^{2}\right)^{1/2}\left(E|\Delta^{N}H^{N}|_{\alpha}^{2}\right)^{1/2}+\\
 & +2\left(E|\Delta^{N}|_{\alpha}^{2}\right)^{1/2}\left(E|p(H_{0}^{N}-H_{0})|_{\alpha}^{2}\right)^{1/2}+\\
 & +2\left(E|\Delta^{N}|_{\alpha}^{2}\right)^{1/2}\left(E|p(V^{N}-V)|_{\alpha}^{2}\right)^{1/2}.
\end{align*}
 Summarizing the results of Lemma $\ref{lem:auxLem}$ we get the final
estimate:
$$
\frac{\mathrm{d}}{\mathrm{d}t}E|\Delta^{N}(t)|_{\alpha}^{2}\leq\mathrm{c}(t)\sup_{s\leq t}E|\Delta^{N}(s)|_{\alpha}^{2}+\frac{\mathrm{c}(t)}{N}.
$$
Grownall's lemma implies the result. It remains to prove Lemma $\ref{lem:auxLem}.$

$\textbf{1.}$ The equation ($\ref{mEq}$) implies $L'(t)\leq C_{\lambda}$, 
therefore $L(t)\leq L(0)+C_{\lambda}t$ which implies $e^{L(t)}\leq\mathrm{c}(t).$
The remaining  inequalities of the first point can be verified in
the same way. 

$\quad\,\,\,\textbf{2.}$ Let $\xi(t)=e^{-L(t)}-e^{-L^{N}(t)},$ then
($\ref{mEq}$) and ($\ref{eq:ZN1}$) imply 
$$
\xi'(t)+C_{\lambda}\xi(t)=\sum_{n\in\mathbb{Z}}(p_{n}(t)-p_{n}^{N}(t))e^{-n}=\sum_{n\in\mathbb{Z}}\Delta_{n}^{N}(t)e^{-n}.
$$
This ODE can be solved explicitly and we get 
\begin{align*}
 & |e^{-L(t)}-e^{-L^{N}(t)}|=|e^{-C_{\lambda}t}\int_{0}^{t}e^{C_{\lambda}s}\sum_{n\in\mathbb{Z}}\Delta_{n}^{N}(s)e^{-n}\mathrm{d}s|\leq c(t)\Gamma(t).
\end{align*}
Therefore using the first statement 
\begin{align*}
 & |e^{L(t)}-e^{L^{N}(t)}|=e^{L(t)+L_{N}(t)}|e^{-L(t)}-e^{-L_{N}(t)}|\leq c(t)\Gamma(t).
\end{align*}
The other statements of the second point can be obtained similarly. 
\begin{rem}
Note that for each $s\in[0,t]$ $p_{n}^{N}(s)$,  has a finite domain
and due to the conditions of the theorem $\ref{thm:existence}$ $p_{n}(t)\lesssim e^{-n^{2}/2+\alpha|n|}$
the value of $\Gamma(t)$ is well-defined.
\end{rem}
$\quad\,\,\,\textbf{3.}$ Applying Cauchy-Schwarz twice we get the third statement
\begin{align*}
 & E\Gamma^{2}(t)=E\left\{ \int_{0}^{t}\sum_{n\in\mathbb{Z}}|\Delta_{n}^{N}(s)|e^{|n|}\mathrm{d}s\right\} ^{2}\leq\\
 & \leq t\int_{0}^{t}E\left\{ \sum_{n\in\mathbb{Z}}|\Delta_{n}^{N}(s)|e^{|n|}\right\} ^{2}\mathrm{d}s=\\
 & =t\int_{0}^{t}E\left\{ \sum_{n\in\mathbb{Z}}\left(\alpha_{n}^{-1/2}\right)\times\left(\alpha_{n}^{1/2}\cdot|\Delta_{n}^{N}(s)|\right)\right\} ^{2}\mathrm{d}s\leq\\
 & \leq t\int_{0}^{t}E\left(\sum_{n\in\mathbb{Z}}\alpha_{n}^{-1}\right)\times\left(\sum_{n\in\mathbb{Z}}\alpha_{n}\cdot|\Delta_{n}^{N}(s)|^{2}\right)\mathrm{d}s=\\
 & =t\left(\sum_{n\in\mathbb{Z}}\alpha_{n}^{-1}\right)\int_{0}^{t}E|\Delta^{N}(s)|_{\alpha}^{2}\mathrm{d}s\leq\mathrm{c}(t)\cdot\sup_{s\leq t}E|\Delta^{N}(s)|_{\alpha}^{2}.
\end{align*}
$\quad\,\,\,\textbf{4.}$ Since $H_{0}^{N}(t)-H_{0}(t)$ is a diagonal
matrix,  
\begin{align*}
 & \left(p(t)(H_{0}^{N}(t)-H_{0}(t)\right)_{n}=\\
 & =p_{n}(t)\left(\lambda_{n}(t)+\mu_{n}(t)-\lambda_{n}^{N}(t)-\mu_{n}^{N}(t)\right)=\\
 & =p_{n}(t)\left\{ e^{-n}\left(e^{L(t)}-e^{L^{N}(t)}\right)+e^{n}\left(e^{-M(t)}-e^{-M^{N}(t)}\right)\right\} .
\end{align*}
Therefore the inequalities of the second statement imply
\begin{align*}
 & |\left(p(t)(H_{0}^{N}(t)-H_{0}(t)\right)_{n}|\leq e^{|n|}\mathrm{c}(t)p_{n}(t)\Gamma(t).
\end{align*}
Now we can prove the fourth statement in the following way:

\begin{align*}
 & E|p(t)(H_{0}^{N}(t)-H_{0}(t))|_{\alpha}^{2}=E\sum_{n\in\mathbb{Z}}\alpha_{n}|\left(p(t)(H_{0}^{N}(t)-H_{0}(t)\right)_{n}|^{2}\leq\\
 & \leq\mathrm{c}(t)E\sum_{n\in\mathbb{Z}}\alpha_{n}e^{|n|}p_{n}^{2}(t)\Gamma^{2}(t)=\mathrm{c}(t)\left(\sum_{n\in\mathbb{Z}}\alpha_{n}e^{|n|}p_{n}^{2}(t)\right)\cdot E\Gamma^{2}(t)\leq\\
 & \leq\mathrm{c}(t)\sup_{s\leq t}E|\Delta^{N}(s)|_{\alpha}^{2}.
\end{align*}
Here in the last inequality we have used the facts that the series $\sum_{n\in\mathbb{Z}}\alpha_{n}e^{|n|}p_{n}^{2}(t)$
converges for each $t$ and is bounded by some nondecreasing function
$\mathrm{c}(t).$ In order to prove this we notice that$\|p(t)\|_{\alpha}\leq\mathrm{c}(t)$
(in  accordance with   theorem $\ref{thm:existence}$) and therefore
$|p_{n}(t)|\leq\mathrm{c}(t)\alpha_{n}^{-1}.$ So 
\begin{align*}
 & \sum_{n\in\mathbb{Z}}\alpha_{n}e^{|n|}p_{n}^{2}(t)\leq\mathrm{c}(t)\sum_{n\in\mathbb{Z}}\alpha_{n}^{-1}e^{|n|}\leq\mathrm{c}(t).
\end{align*}

$\quad\,\,\,\textbf{5.}$ The inequalities of the second statement
imply 
\begin{align*}
 & |\left(p(t)(V(t)-V^{N}(t))\right)_{n}|=\\
 & =|e^{-n+1}p_{n-1}(t)(e^{L(t)}-e^{L_{N}(t)})+e^{n+1}p_{n+1}(t)(e^{-M(t)}-e^{-M_{N}(t)})|\leq\\
 & \leq\mathrm{c}(t)\Gamma(t)\left(e^{-n+1}p_{n-1}(t)+e^{n+1}p_{n+1}(t)\right).
\end{align*}
Therefore,  
\begin{align*}
 & E|p(t)(V^{N}(t)-V(t))|_{\alpha}^{2}=\\
 & =\mathrm{c}(t)E\sum_{n\in\mathbb{Z}}\alpha_{n}\Gamma^{2}(t)\left(e^{-n+1}p_{n-1}(t)+e^{n+1}p_{n+1}(t)\right)^{2}\leq\\
 & \leq\mathrm{c}(t)E\sum_{n\in\mathbb{Z}}\alpha_{n}\Gamma^{2}(t)\left(e^{-2n+2}p_{n-1}^{2}(t)+e^{2n+2}p_{n+1}^{2}(t)\right)=\\
 & =\mathrm{c}(t)E\sum_{n\in\mathbb{Z}}\Gamma^{2}(t)p_{n}^{2}(t)\left(\alpha_{n+1}e^{-2n}+\alpha_{n-1}e^{2n}\right)=\\
 & =\mathrm{c}(t)E\Gamma^{2}(t)\times\sum_{n\in\mathbb{Z}}p_{n}^{2}(t)\left(\alpha_{n+1}e^{-2n}+\alpha_{n-1}e^{2n}\right)\leq\\
 & \leq\mathrm{c}(t)E\Gamma^{2}(t)\leq\mathrm{c}(t)\sup_{s\leq t}E|\Delta^{N}(s)|_{\alpha}^{2}.
\end{align*}
Here in the second row we have used the inequality $(a+b)^{2}\leq2a^{2}+2b^{2}.$
In the third row we shifted the indecies of summation and in the fifth
row we have used that 
$$
\sum_{n\in\mathbb{Z}}p_{n}^{2}(t)\left(\alpha_{n+1}e^{-2n}+\alpha_{n-1}e^{2n}\right)\leq\mathrm{c}(t).
$$
Indeed, since $|p_{n}(t)|\leq\mathrm{c}(t)\alpha_{n}^{-1}$ we have
\begin{align*}
 & p_{n}^{2}(t)\left(\alpha_{n+1}e^{-2n}+\alpha_{n-1}e^{2n}\right)\leq\mathrm{c}(t)\alpha_{n}^{-2}\left(\alpha_{n+1}e^{-2n}+\alpha_{n-1}e^{2n}\right)\leq\mathrm{c}(t)e^{-\frac{n^{2}}{2}+(\alpha+3)|n|}
\end{align*}
and therefore the corresponding series converges.

$\quad\,\,\,\textbf{6.}$ We have 
\begin{align*}
 & \Delta_{n}^{N}\left(\Delta^{N}H^{N}\right)_{n}=\\
 & =\Delta_{n}^{N}\left(\Delta_{n-1}^{N}\lambda_{n-1}^{N}+\Delta_{n+1}^{N}\mu_{n+1}^{N}-\Delta_{n}^{N}\lambda_{n}^{N}-\Delta_{n}^{N}\mu_{n}^{N}\right)=\\
 & =\Delta_{n}^{N}\left(\Delta_{n-1}^{N}e^{-n+1+L^{N}}+\Delta_{n+1}^{N}e^{n+1-M^{N}}-\Delta_{n}^{N}e^{-n+L^{N}}-\Delta_{n}^{N}e^{n-M^{N}}\right)=\\
 & =e^{-n+L^{N}}\left(e\Delta_{n-1}^{N}\Delta_{n}^{N}-(\Delta_{n}^{N})^{2}\right)+e^{n-M^{N}}\left(e\Delta_{n}^{N}\Delta_{n+1}^{N}-(\Delta_{n}^{N})^{2}\right)\leq\\
 & \leq\mathrm{c}(t)\left\{ e^{-n}\left(e\Delta_{n-1}^{N}\Delta_{n}^{N}-(\Delta_{n}^{N})^{2}\right)+e^{n}\left(e\Delta_{n}^{N}\Delta_{n+1}^{N}-(\Delta_{n}^{N})^{2}\right)\right\} .
\end{align*}
Here in the last inequality we have used the first statement. Now
we use the following simple inequality 
$$
eab-b^{2}=(2^{-1/2}ea)\times(2^{1/2}b)-b^{2}\leq e^{2}a^{2}/4+b^{2}-b^{2}=e^{2}a^{2}/4
$$
and get 
\begin{align*}
 & \Delta_{n}^{N}\left(\Delta^{N}H^{N}\right)_{n}\leq\mathrm{c}(t)\left(e^{-n}\left(\Delta_{n-1}^{N}\right)^{2}+e^{n}\left(\Delta_{n+1}^{N}\right)^{2}\right).
\end{align*}
Therefore,  
\begin{align*}
 & E\langle\Delta^{N},\Delta^{N}H^{N}\rangle_{\alpha}=\\
 & =E\sum_{n\in\mathbb{Z}}\alpha_{n}\Delta_{n}^{N}\left(\Delta^{N}H^{N}\right)_{n}\leq\\
 & \leq\mathrm{c}(t)E\sum_{n\in\mathbb{Z}}\alpha_{n}\left(e^{-n}(\Delta_{n-1}^{N})^{2}+e^{n}(\Delta_{n+1}^{N})^{2}\right)=\\
 & =\mathrm{c}(t)E\sum_{n\in\mathbb{Z}}(\Delta_{n}^{N})^{2}\left(\alpha_{n+1}e^{-n-1}+\alpha_{n-1}e^{n-1}\right)\leq\\
 & \leq\mathrm{c}(t)E\sum_{n\in\mathbb{Z}}\alpha_{n}(\Delta_{n}^{N})^{2}=\mathrm{c}(t)|\Delta^{N}|_{\alpha}^{2}.
\end{align*}
Here in the last inequality we have used ($\ref{eq:aux1}$) and ($\ref{eq:aux2}$).

$\quad\,\,\,\textbf{7.}$ First note that the inequalities of the
first statement imply 
\begin{align*}
 & Q_{n}^{N}(t)\leq\mathrm{c}(t)\left(e^{-n}p_{n-1}^{N}(t)+e^{n}p_{n+1}^{N}(t)+p_{n}^{N}(t)e^{-n}+p_{n}^{N}(t)e^{n}\right).
\end{align*}
Therefore,  
\begin{align*}
 & \sum_{n\in\mathbb{Z}}\alpha_{n}Q_{n}^{N}(t)\leq\\
 & \leq\mathrm{c}(t)\sum_{n\in\mathbb{Z}}\alpha_{n}\left(e^{-n}p_{n-1}^{N}(t)+e^{n}p_{n+1}^{N}(t)+p_{n}^{N}(t)e^{-n}+p_{n}^{N}(t)e^{n}\right)\leq\\
 & \leq\mathrm{c}(t)\sum_{n\in\mathbb{Z}}p_{n}^{N}(t)\left(e^{-n-1}\alpha_{n+1}+e^{n-1}\alpha_{n-1}+e^{-n}\alpha_{n}+e^{n}\alpha_{n}\right)\leq\\
 & \leq\mathrm{c}(t)\sum_{n\in\mathbb{Z}}\alpha_{n}e^{|n|}p_{n}^{N}(t)\leq\mathrm{c}(t)\|p^{N}(t)\|_{\alpha+1}.
\end{align*}
Here in the third inequality we have used ($\ref{eq:aux1}$) and ($\ref{eq:aux2}$)
and in the last we used ($\ref{eq:aux3}$). As a consequence,
\begin{align*}
 & E\sum_{n\in\mathbb{Z}}\alpha_{n}Q_{n}^{N}(t)\leq\mathrm{c}(t)\|p_{E}^{N}(t)\|_{\alpha+1}, 
\end{align*}
where $p_{E}^{N}(t)=Ep_{n}^{N}(t)=P(x_{i}^{N}(t)=n),$ $i\in\left\{ 1,\ldots,N\right\} .$
Consider an auxiliary Markov chain $\tilde{x}(t)$ on $\mathbb{Z}$
with the jumps of the following intensities 
\begin{align*}
n\to n+1: & \tilde{\lambda}_{n}(t)=\exp\left\{ -|n|+\max\left(+L(0),-M(0)\right)+C_{\lambda}t\right\} \cdot I\left\{ n\geq0\right\},  \\
n\to n-1: & \tilde{\mu}_{n}(t)=\exp\left\{ -|n|+\max\left(+L(0),-M(0)\right)+C_{\mu}t\right\} \cdot I\left\{ n<0\right\} .
\end{align*}

\begin{rem}
$\tilde{x}(t)$ jumps only to the right or only to the left.
\end{rem}
Denote by $\tilde{p}(t)=\left\{ \tilde{p}_{n}(t)\right\} $ its distribution
at the moment $t\in[0,\infty).$ By analogy with lemma 2.4 in [19]
it is easy to prove that $\tilde{x}(t)$ corresponds to a strongly continuous
propagator $\tilde{P}(s,t),$ $0\leq s\leq t<\infty$ on $B_{\alpha+1}$
such that $\tilde{p}(t)=\tilde{p}(s)\tilde{P}(s,t).$ As a consequence
$$
\|\tilde{p}(t)\|_{\alpha+1}\leq\|p(0)\|_{\alpha+1}\|\tilde{P}(0,t)\|_{\alpha+1}=\mathrm{c}(t)\|p(0)\|_{\alpha+1}.
$$
Notice that between $x_{i}^{N}(t)$ and $\tilde{x}(t)$ we can construct
a coupling such that $|x_{i}^{N}(t)|\geq|\tilde{x}(t)|$. As a consequence, 
($\ref{eq:aux4}$) implies 
$$
\|p_{E}^{N}(t)\|_{\alpha+1}=Ee^{\frac{cx_{i}^{N}(t)^{2}}{2}+\alpha|x_{i}^{N}(t)|}\leq Ee^{\frac{c\tilde{x}(t)^{2}}{2}+\alpha|\tilde{x}(t)|}\leq\|\tilde{p}(t)\|_{\alpha+1}, 
$$
and we get the desired result.

\end{document}